\documentclass{iau} 
\usepackage{graphicx, color}

\title[Pulsar Searches with the SKA] 
{Pulsar Searches with the SKA}

\author[Lina Levin \& the SKA-TDT team]   
{L. Levin$^{1,*}$, 
W. Armour$^2$, C. Baffa$^3$, E. Barr$^4$, S. Cooper$^1$, \\ 
R. Eatough$^4$, A. Ensor$^5$, E. Giani$^3$, A. Karastergiou$^2$, \\
R. Karuppusamy$^4$, M. Keith$^1$, M. Kramer$^4$, R. Lyon$^1$, \\
 M. Mackintosh$^6$, M. Mickaliger$^1$, R van Nieuwpoort$^7$, \\
M. Pearson$^1$, T. Prabu$^1$, J. Roy$^{1,8}$, O. Sinnen$^9$, L. Spitler$^4$, \\
H. Spreeuw$^7$,  B. W. Stappers$^1$, W. van Straten$^5$, C. Williams$^2$,\\
 H. Wang$^9$,  K. Wiesner$^4$, 
  \and the SKA TDT team}

\affiliation{$^1$Jodrell Bank Centre for Astrophysics, University of Manchester,\\ Oxford Road, M13 9PL, Manchester, UK\\[\affilskip]
$^2$University of Oxford, Denys Wilkinson Building, Keble Road, Oxford OX1 3RH, UK\\[\affilskip]
$^3$INAF-Osservatorio Astrofisico di Arcetri, Largo E. Fermi 5, 50125, Firenze, Italy\\[\affilskip]
$^4$Max-Planck-Institut fur Radioastronomie, Auf dem H\"{u}gel 69, D-53121 Bonn, Germany\\[\affilskip]
$^5$Institute for Radio Astronomy \& Space Research, Auckland University of Technology,\\ Private Bag 92006, Auckland 1142, New Zealand\\[\affilskip]
$^6$Science and Technology Facilities Council,\\ Polaris House, North Star Avenue, Swindon, SN2 1SZ, UK\\[\affilskip]
$^7$ Netherlands Institute for Radio Astronomy (ASTRON),\\ Postbus 2, NL-7990 AA Dwingeloo, the Netherlands\\[\affilskip]
$^8$ NCRA-TIFR, Pune University Campus, Pune - 411007, India\\[\affilskip]
$^9$ Parallel and Reconfigurable Computing (PARC) lab, University of Auckland,\\ Private Bag 92019, Auckland 1142, New Zealand\\[\affilskip]
$^*$email: {\tt lina.preston@manchester.ac.uk}}

\pubyear{2017}
\volume{337}  
\jname{Pulsar Astrophysics -­ The Next 50 Years}
\editors{P. Weltevrede, B.B.P. Perera, L. Levin Preston \& S. Sanidas, eds.}

\begin{document}

\maketitle

\begin{abstract}
The Square Kilometre Array will be an amazing instrument for pulsar astronomy. 
While the full SKA will be sensitive enough to detect all pulsars in the Galaxy visible from Earth, already with SKA1, pulsar searches will discover enough pulsars to increase the currently known population by a factor of four, no doubt including a range of amazing unknown sources. 
Real time processing is needed to deal with the 60 PB of pulsar search data collected per day, using a signal processing pipeline required to perform more than 10 POps. 
Here we present the suggested design of the pulsar search engine for the SKA and discuss challenges and solutions to the pulsar search venture.
\end{abstract}

The Square Kilometre Array (SKA) will be an excellent telescope for discovering and timing pulsars (see e.g. \cite[Keane et al., 2015; Baffa 2014;]{kea15,baf14} and contribution by E. Keane in these proceedings). Already the first stage of the telescope (SKA1) will more than quadruple the known population of pulsars, and once the full SKA is finished, it will be sensitive enough to find all pulsars in the Galaxy that are beaming towards Earth. 
Many of these will be exciting sources, encompassing all different types of pulsars, such as millisecond pulsars (MSPs), rotating radio transients, young pulsars, magnetars and pulsars in binary systems. In particular, the SKA will have excellent sensitivity to highly relativistic double neutron star systems as well as pulsars orbiting black holes. 
To enable the discovery of all these sources, it is essential to have a fast and stable search pipeline installed at the telescope. Here we outline our proposed search strategy, address the challenges involved in performing such a search, as well as present our prototyping efforts aimed at testing and verifying the pulsar search hardware and software.

\section{Pulsar Search Parameters}
Pulsar science is one of the SKA Key Science Drivers, and pulsar searches will be performed on both SKA1-Low and SKA1-Mid. On SKA1-Low, 500 tied-array beams will be formed, producing pulsar search data over a $\sim$100-MHz band sampled every $\sim$100\,$\mu$s, while 1500 tied-array beams will be formed on SKA1-Mid, with data collected over a $\sim$300-MHz band sampled every $\sim$64\,$\mu$s. Current parameters relating to the pulsar search engine are summarised in Table \ref{tab1}. 

Simulations using these parameters show that a combined all-sky pulsar survey, where SKA1-Mid is focussing on the Galactic plane ($|b|<10^\circ$ and $\delta<45^\circ$) and SKA1-Low is covering the higher Galactic latitudes ($|b|>5^\circ$ and $\delta<30^\circ$), will discover $\sim$9000 ordinary pulsars and $\sim$1200 MSPs.

\begin{table}
  \begin{center}
  \caption{Pulsar search parameters for SKA1.}
  \label{tab1}
 {\scriptsize
  \begin{tabular}{l l l }\hline 
{\bf Parameter} 					& {\bf SKA1-Mid} 		& {\bf SKA1-Low} \\ \hline
Frequency bands [MHz]			& Band 1: 350 - 1050 	& 50 - 350 \\
							& Band 2: 950 - 1760 	&  \\
							& Band 5: 4600 - 13800 	&  \\
Simultaneous beams 			& 1500 				& 500 \\
Bandwidth [MHz] 				& $\sim$300			& $\sim$100 \\ 
Frequency channels				& 4096				& 8192\\
Time resolution [$\mu$s]			& $\sim$64 			& $\sim$100 \\
Polarisations 					& 4 					& 4 \\
Bits per sample					& 8					& 8 \\
  \end{tabular}
  }
 \end{center}
\end{table}

\section{The SKA Pulsar Search Engine}
The combination of number of beams, sampling rate and number of frequency channels for the SKA pulsar searches result in very high incoming data rates. As an example, assuming a survey with SKA1-Mid using Band\,2 results in data reaching the pulsar search engine at a rate of $\sim$800\,GB/s, and one full day of observing would collect $\sim$60\,PB of data. With such high data rates, it is necessary to perform the pulsar search in real time. This is an intricate task, especially considering the many complex algorithms involved in performing pulsar acceleration searches. 
Building a telescope of the size of the SKA also involves a lot of additional requirements, and another important consideration when building the data processing machines is to fit within the very limited power budget. 
To address these challenges and create a pulsar search engine for the SKA that fulfils all the mentioned requirements, we have designed a dedicated compute cluster as part of the SKA Central Signal Processor (CSP). In our design, this cluster will consist of 500 compute nodes at SKA1-Mid and 167 nodes at SKA1-Low, each able to process up to three pulsar search beams in real time. To achieve the required performance, each node will include two accelerators, which currently are proposed to be one GPU card and one FPGA board. This combination of GPUs and FPGAs will allow us to take advantage of the great speed of parallel processing in the GPUs as well as the very power-efficient processing of the FPGAs. It gives us flexibility to use the processor best suited for a particular task, for example the GPUs could be used for dedispersion and single pulse detection, while the FPGAs could be used for the Fourier domain acceleration processing.

\subsection{Pulsar Processing on the Central Signal Processor}
\begin{figure}[b]
\begin{center}
 \includegraphics[width=5in]{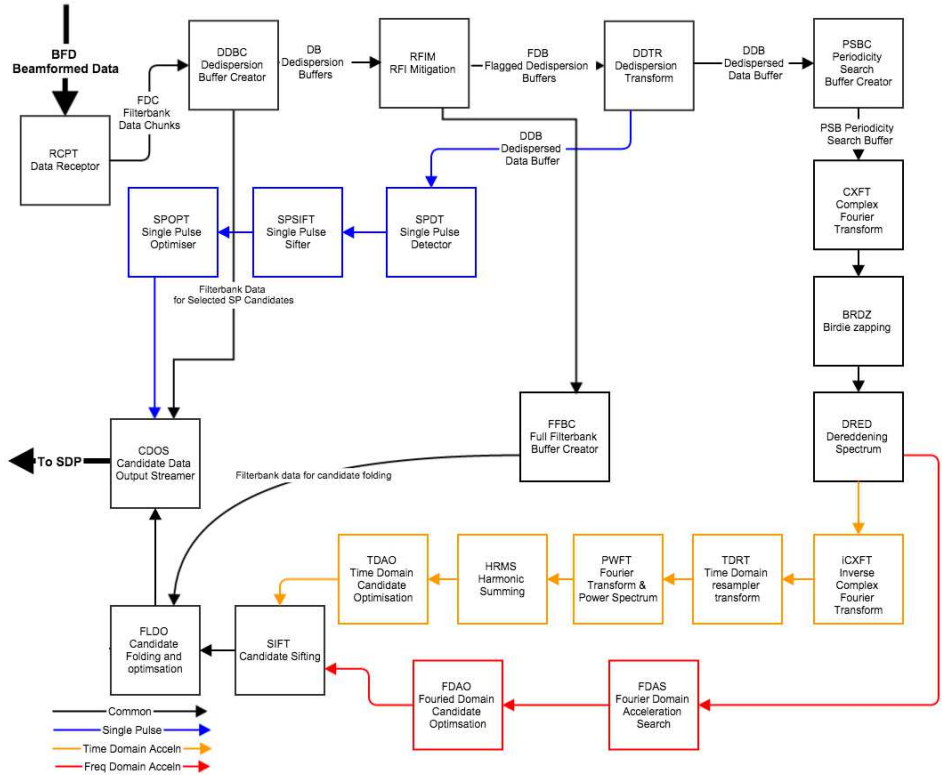} 
 \caption{Schematic of the pulsar search processing work flow within CSP.}
   \label{fig1}
\end{center}
\end{figure}

As is conventional for pulsar searches, each beam-formed data stream will be searched in dispersion measure (DM) space, to account for delays over the observed frequency band caused by the free electrons in the interstellar medium. This search will be performed on up to 6000 DM trials, which requires $\sim$2\,TOps of computing, assuming an integration time of $\sim$10\,minutes. Thereafter each of the resulting DM-corrected timeseries will go through a Fast Fourier Transform (FFT). The length of the FFT will depend on the observation time, but in the example of a $\sim$10-minute integration, this results in an 8-million point FFT of each of the 6000 timeseries, adding up to a total of $\sim$1\,TOp of computing. 
To search for pulsars in relativistic binary systems, it is also necessary to account for changes in pulse period caused by binary motion. We carry out this acceleration search in the Fourier domain, similar to \cite{ran02}, but are also testing time domain acceleration search techniques. In the Fourier domain search, up to 100 acceleration trials are performed on at least 500 DM-corrected timeseries, which requires $\sim$10\,POps of computing. The acceleration search produces lists of pulsar candidate signals for each DM and acceleration trial. These need to be compared and sifted to account for multiple detections of the same signal in different trials, as well as detection of multiple harmonics of the fundamental pulsar signal.  
Finally, up to 1000 pulsar candidate files will be created by folding the data at the resulting pulse period, DM and acceleration, requiring $\sim$140\,TOps. 
An overview of the pulsar search pipeline within CSP is shown in Fig. \ref{fig1}.

\subsection{Machine Learning on the Science Data Processor}
The dedicated cluster at CSP will process the pulsar search data and produce pulsar candidate files. These will need to be sifted and ranked to sort out the real pulsars from the interference and noise. All pulsar candidate files from each search beam will be sent from CSP to the processing cluster at the SKA Science Data Processor (SDP). When arriving at SDP the candidate data from all the beams will be compared in a so-called multi-beam sift. This will allow us to identify interference signals present in a large number of beams simultaneously, as well as to choose the most significant detection of a pulsar, if it has been detected in multiple beams and/or as multiple harmonics of the pulsar signal. The candidates that survive this multi-beam sift will be passed on to a candidate classification routine which will extract heuristic features from each candidate and use these to select the most likely real pulsars using machine learning techniques (\cite[e.g. Lyon et al., 2016; 2017]{lyo16, lyo17}). Once a pulsar candidate has been selected as a real pulsar, it will be marked for follow-up at the telescope. All data for the candidates that pass the multi-beam test, as well as all meta-data and heuristic scores for all candidates coming from CSP, will be saved long term in the SKA archive.

\section{ProtoNIP}
To test our newly developed algorithms and trial the proposed CSP pulsar processing nodes, we have constructed ProtoNIP, a prototype processing cluster. ProtoNIP is a 3\% prototype of the full CSP pulsar machine, located on the MeerKAT site in South Africa. By placing it on the site that will eventually be the location of SKA-Mid, we are in a perfect position to test how the exact system that will be implemented on the final SKA functions in the right environment. This will enable us to make accurate predictions on important aspects such as power usage, cooling systems, processing pipeline timings and system stability. 

In addition, we have set up a test vector machine at Jodrell Bank Observatory, which is directly interfaced with a ProtoNIP node. This machine allows us to run tests on each individual module of the search pipeline, using test vectors with the same dimensions as are expected for the SKA searches. We have created test vectors for a wide range of simulated pulsars with different properties, and these will be used to test both the individual modules and the combined search pipeline. 
The test vector machine will also be used to further develop heuristic features and classification models for the candidate selection tasks. Machine learning will be run on both simulated and real data containing a wide range of pulsars to assure we get the highest return possible of newly discovered pulsars. 

The knowledge gained from ProtoNIP and the test vector analysis will be vitally important in the development of the full processing clusters, and ensure the success of pulsar searching with the SKA.\\

{}  

\end{document}